\documentclass[epsfig]{emulateapj}

\def\logh{5\,\mbox{log}\,h}
\newcommand{\petroratio}{{{\mathcal{R}}_P}}
\newcommand{\petroradius}{{{\theta}_P}}

\slugcomment{Accepted in AJ}
\shorttitle{Dispersion Around the Color-Magnitude Relation}
\shortauthors{Cool et al.}

\begin{document}
\title{Broadband Optical Properties of Massive Galaxies:
 the Dispersion Around the Field Galaxy Color-Magnitude Relation
 Out to $z\sim0.4$}
\author{Richard J. Cool\altaffilmark{1}, Daniel J. Eisenstein\altaffilmark{1}, 
David Johnston\altaffilmark{2}, Ryan Scranton\altaffilmark{3}, 
Jon Brinkmann\altaffilmark{4},   Donald P. Schneider\altaffilmark{5}, 
Idit Zehavi\altaffilmark{1} }

\altaffiltext{1}{Steward Observatory, 933 N Cherry Avenue, Tucson AZ 85721; 
rcool@as.arizona.edu}
\altaffiltext{2}{Department of Astrophysical Sciences, Peyton Hall,
 Princeton University, Princeton, NH 08544}
\altaffiltext{3}{University of Pittsburgh, Department of Physics and 
Astronomy, 3941 O'Hara Street, Pittsburgh, PA 15260}
\altaffiltext{4}{Apache Point Observatory, P.O. Box 59, Sunspot, NM 88349}
\altaffiltext{5}{Department of Astronomy, The Pennsylvania State 
University, University Park, PA 16802}

\begin{abstract} 
Using a sample of nearly 20,000 massive early-type galaxies selected from the 
Sloan Digital Sky Survey, we study the color-magnitude relation for the most 
luminous ($L \gtrsim 2.2 L^{*}$) field galaxies in the redshift range 
$0.1<z<0.4$ in several colors.  The intrinsic dispersion in galaxy colors is 
quite small in all colors studied, but the 40 milli-mag scatter in the bluest 
colors is a factor of two larger than the 20 milli-mag measured in the reddest 
bands. While each of three simple models constructed for the star formation 
history in these systems can satisfy the constraints placed by our
measurements, none of them produce color distributions matching those 
observed.  Subdividing by environment, we find the dispersion for galaxies 
in clusters to be  about 11\% smaller than that of more isolated systems. 
Finally, having resolved the red sequence, we study the color dependence 
of the composite spectra. Bluer galaxies on the red sequence are found to 
have more young stars than red galaxies; the extent of this spectral 
difference is marginally better described 
by passive evolution of an old stellar
population than by a model consisting of a recent trace injection of 
young stars.

\end{abstract}

\section{Introduction}
The presence of a tight correlation between the rest-frame optical color 
and luminosity of early-type galaxies, the so-called color-magnitude
relation (CMR) or red-sequence, is well established in the literature 
\citep{vs1977,ltc1980, lugger1984, zepf1991, 
 terlevich1999, blanton2003a1,hogg2004, baldry2004, lby2004}.  
The relation implies that the properties of the 
stellar populations in early-type galaxies are strongly dependent on the
total stellar mass of the system.  The tight correlation observed along 
this relationship \citep[e.g.][]{faber1973, vs1977, ble1992} requires a 
strong coherence between both the age and metallicity of stellar populations 
present in early-type galaxies.  Measurements of the slope, zero-point, and
dispersion around the color-magnitude relation probe the star formation
histories of
early-type galaxies and can provide physical insight into galaxy formation
and evolution.

One possible explanation for the slope of the CMR is that massive, and 
thus more luminous, galaxies retain more metals than less massive ones.  
The deeper potential wells of massive galaxies may allow fewer metals to 
be expelled in supernova driven winds \citep{larson1974, ay1987, mt1987,
bct1996}.  Alternatively, \citet{kc1998} have shown that the color-magnitude
relation can be reproduced in hierarchical merging models if the epoch of
mergers occurs at very early times. Since the age of a stellar population
is degenerate with its metallicity on the CMR, however, metallicity may 
not be the sole parameter behind the relationship \citep{faber1972,faber1973,
oconnell1980,rose1985,wtf1995}. While both age and metallicity effects may
create the CMR, studies of the evolution of the relationship toward
high-redshift argue that the relation is driven primarily by metallicity
and cannot be generated by age effects alone \citep{ka1997}.  

The color-magnitude relation of galaxies in clusters has been studied
extensively.  Using high precision $U$ and $V$ photometry, \citet{ble1992}
determined the dispersion around the CMR for early-type galaxies in Coma
and Virgo to be small, $\delta(U-V)\lesssim 40$ milli-mag (mmag). 
This small scatter
implies early-type galaxies in clusters must be uniformly old with
formation epochs earlier than a $z\sim2$.  \citet{ellis1997} found 
$\delta(U-v)\sim 100$ mmag for clusters at $z\approx0.54$; this scatter, 
combined with the early age of the Universe at the observed epoch, indicates
that elliptical galaxies must have stopped forming stars in these clusters
by a $z\sim2.7$ for a simple approximation to the color evolution of 
stellar populations. The cluster MS 1054-03 at a $z=0.83$  has a 29 mmag
dispersion in the $U-B$ color 
\citep{vandokkum2000}, while \citet{vandokkum2001}
and \citet{blakeslee2003} showed that the scatter in single clusters is 
only 40 and 30 mmag at $z\sim1.3$. These small dispersions, seemingly 
independent of redshift, indicate that the coordination of star formation
between galaxies in clusters is quite strong.   Using a sample of 158 
clusters with $0.06<z<0.34$ drawn from the Early Data Release of the Sloan
Digital Sky Survey (SDSS), \citet{andreon2003} found that the CMR between 
clusters is very homogeneous. Locally, the CMR is universal between clusters
of various masses as well \citep{mzr2004}.  

While the mean color of red sequence galaxies in over-dense regions is only 
slightly redder than for similar galaxies in less dense regions and the slope
of the CMR is independent of environment \citep{hogg2004}, several studies 
show that the scatter around the CMR is dependent on galaxy environment.  
A re-analysis of the data presented in \citet{sv1978} indicates the
dispersion around the CMR in clusters is 2-$\sigma$ smaller than that 
for galaxies in groups or in the field \citep{ltc1980}.  The scatter in 
galaxy colors in the core of the cluster CL 1358+62 at $z=0.33$ have 
$\delta(B-V)\sim 22$ mmag while early-types at large radii ($R>1.4h^{-1}$ Mpc) 
have nearly double that value, indicating that the dispersion of galaxy 
colors in the field may be larger than that measured in clusters 
\citep{vandokkum1998}. Early-type galaxies in the Hubble Deep Field 
have a scatter of 120 $\pm$ 60 mmag in rest-frame $U-V$ with an average 
redshift of 0.9 \citep{kbb1999}.  While this measurement is rather uncertain, 
it is larger than that measured in clusters, further indicating that field 
galaxies may have less coordinated star formation than galaxies in clusters. 

A majority of the work on the CMR has focused on early-type galaxies in 
clustered environments with luminosities near $L^*$.  This is natural, as 
statistically significant samples are more easily selected in these 
over-dense regions and $L^*$ galaxies are more common than their more 
massive analogs.  In this work, we extend the analysis of the color-magnitude
relation to a sample of nearly 20,000 very luminous red field galaxies from 
the Sloan Digital Sky Survey.  These galaxies are not selected to reside 
in clustered environments and are chosen to be quite luminous with 
$L \gtrsim 2.2 \, L^{*}$.   We create two galaxy samples from the SDSS 
spectroscopic data. At low-redshifts ($0.1<z<0.2$) we use galaxies from 
the entire survey region while at higher redshifts ($0.3<z<0.4$), where 
the galaxies are apparently fainter, we select galaxies from a 270 deg$^2$
region which has been imaged an average of 10 times and up to 29 times.
The repeated imaging of our moderate-redshift galaxies makes comparisons
between the galaxy samples possible with similar fidelity. Our sample
allows us to probe a new area in parameter space --- luminous galaxies 
in the field --- and compare our results to those found for galaxies in
clustered environments.

The paper is organized as follows: \S2  describes the Sloan Digital Sky 
Survey and the selection for the galaxies used here. In \S 3, we present
our measurements of the slope and scatter of the  color-magnitude relations. 
We discuss simple star formation history models in \S 4 and use our
measurements of the scatter around the CMR as a constraint on the 
evolutionary history of massive early-type galaxies in the field.  
In \S 5, we discuss the role environment plays on the CMR of galaxies.
We construct the composite spectra of galaxies across the red sequence
in \S 6 before closing in \S 7.  Throughout this work,  we use $H_{0}$
= $100\,h$ km s$^{-1}$ Mpc$^{-1}$ and ($\Omega_{\rm m}$, $\Omega_{\Lambda}$)
= (0.3,0.7) to calculate luminosities and distances. When calculating
ages of stellar populations, we set $h=0.7$.  All magnitudes used here
have been corrected for 
reddening using the \citet{sfd1998} dust maps.

\pagebreak
\section{Data}
\subsection{The Sloan Digital Sky Survey}

The Sloan Digital Sky Survey  \citep{york2000, stoughton2002b,a2003,a2004a,
a2004b} is imaging $\pi$ steradians of the northern sky through 5 passbands 
\citep{fukugita1996}.  The imaging is conducted with a CCD mosaic in 
drift-scanning mode \citep{gunn1998} on a dedicated 2.5m  telescope located 
at  Apache Point Observatory in New Mexico. After the images are processed 
\citep{lupton2001, stoughton2002, pier2003} and calibrated \citep{hogg2001, 
smith2002, ivezic2004}, targets are selected for spectroscopy with two 
double-spectrographs mounted on the same telescope using an automated 
spectroscopic fiber assignment algorithm \citep{blanton2003a}. 

Two spectroscopic galaxy samples are created using the SDSS imaging. The 
MAIN galaxy sample \citep{strauss2002} is a complete, flux limited, sample 
of galaxies with $r<17.77$. This cut is nearly five magnitudes brighter than 
the SDSS detection limit of $r \sim $ 22.5, and thus the photometric 
quantities for these galaxies are measured with signal-to-noise of a few 
hundred.  The Luminous Red Galaxy (LRG) sample \citep{eisenstein2001} 
selects luminous  early-type galaxies out to $z\sim0.5$ with $r<$ 19.5 
using several color-magnitude cuts in $g$, $r$, and $i$. The combination 
of these two samples allows us to study the broadband colors of luminous 
field galaxies at moderate-redshifts with statistically significant samples.  

While the focal point for the SDSS is a contiguous survey of the Northern 
Galactic Cap, the SDSS also conducts a deep imaging survey, the SDSS Southern 
Survey, by repeatedly imaging an area on the celestial equator in the 
Southern Galactic Cap.  Currently, objects in the 270 deg$^2$ region have 
been imaged an average of 10 times and up to 29 times, resulting in improved 
photometric precision for faint galaxies.  
Objects detected in each observational 
epoch were matched using a tolerance of 0.5 arcseconds to 
create the final coadded catalog used here.
The photometric measurements from each epoch were combined by converting the 
reported asinh magnitudes \citep{LGS1999} into
flux and then calculating the mean value.  
Errors on each parameter are simply the 
standard deviation of the flux measurements.
The Southern Survey is an ideal region to compare 
the properties of faint galaxies at moderate-redshift 
with  brighter low-redshift analogs drawn from the entire SDSS survey area 
with photometry of similar fidelity.

\begin{figure*}[ht]
\centering{\includegraphics[angle=-90, width=6in]{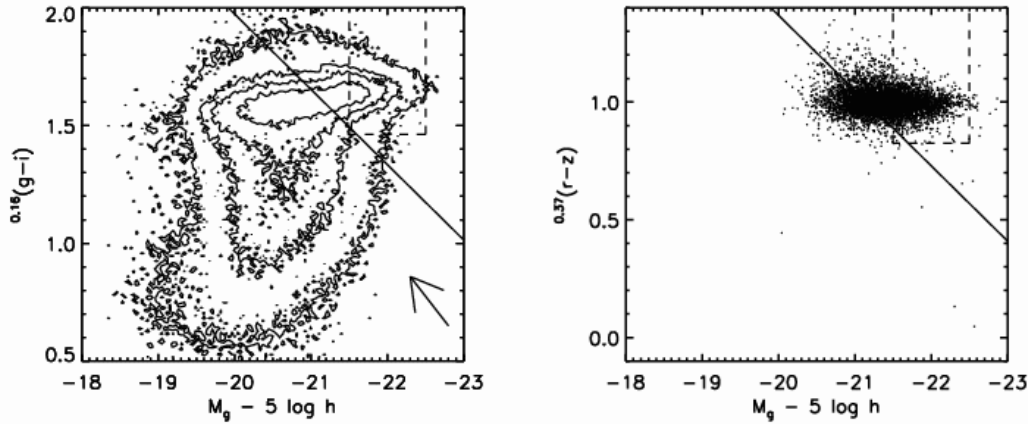}}
\caption{\scriptsize Sample definition for the galaxies used in this paper.
The left panel shows the low redshift selection regions; contours encapsulate
25\%, 50\%, 75\%, and 95\% of the considered galaxies.  The right panels
shows the target selection for our high-redshift galaxies overlayed on 
the LRG galaxies in the SDSS Southern Survey.  In each panel, the solid 
line shows the selection cut which approximates a cut at constant stellar
mass using the relations of \citet{BdJ2000}.  The regions defined by 
the dashed lines were used to create control samples which do not have a
strong cut across the red sequence to test the effects of sample selection
on our scatter measurements. The arrow in the lower right of the left panel
shows the reddening vector.  Note the dearth of very luminous blue galaxies 
which could contaminate our red-sequence samples if heavily reddened.  }
\label{sample}
\end{figure*}

\subsection{Galaxy Photometric Properties}

Several methods are used to measure galaxy fluxes in SDSS.  Below, 
we briefly describe the two flux measurements used throughout this paper.
The Petrosian ratio, 
$\petroratio$, the ratio of the local surface brightness at radius $\theta$ 
to the average surface brightness at that radius, is given by 
\begin{equation}
\label{petroratio}
\petroratio (\theta)\equiv \frac{\left.
\int_{\alpha_{\mathrm{lo}} \theta}^{\alpha_{\mathrm{hi}} \theta} d\theta' 2\pi \theta'
I(\theta') \right/ \left[\pi(\alpha_{\mathrm{hi}}^2 -
\alpha_{\mathrm{lo}}^2) \theta^2\right]}{\left.
\int_0^\theta dr' 2\pi \theta'
I(\theta') \right/ [\pi \theta^2]},
\end{equation}
where $I(\theta)$ is the azimuthally averaged surface brightness profile 
of a galaxy and $\alpha_{\mathrm{lo}}$ and $\alpha_{\mathrm{hi}}$ are chosen to be
0.85 and 1.25 for SDSS. The Petrosian flux is given by the flux within a circular 
aperture of $2 \petroradius$, where $\petroradius$ is the radius at which 
$\petroratio$ falls below 0.2.  In SDSS, $\petroradius$ is determined in 
the $r$ band then subsequently used in each of the other bands.  This flux 
measurement contains a constant fraction of a galaxy's light,  independent 
of its size or distance, in the absence of seeing. More details of SDSS 
Petrosian magnitudes can be found in \citet{blanton2001}, and 
\citet{strauss2002}.

For each galaxy imaged by SDSS, two seeing-convolved models, a pure  \citet{dv1948} profile 
and a pure exponential profile, are fit to the galaxy image. The best-fit 
model in the $r$ band is used to measure the flux of a galaxy in each of 
the other bands. These model magnitudes are unbiased in the absence of color 
gradients and provide higher signal-to-noise ratio colors than Petrosian 
colors.  A more complete description of  model magnitudes is given in 
\citet{stoughton2002b}.  Throughout this paper, we use Petrosian magnitudes 
when calculating galaxy luminosities while model magnitudes are used to 
determine galaxy colors.  

\subsection{The $k$-Correction}

We calculate photometric $k$-corrections  using the method of 
\citet{blanton2003b} (\texttt{kcorrect v3\_2}). This program constructs
a linear combination of carefully chosen spectral templates in order to best 
match the observed photometry at the measured redshift of the galaxy.  The 
rest-frame colors of these best-fit spectra are used to derive corrections 
to the observed galaxy colors.  With this method, it becomes simple to 
transform between bandpasses and, if necessary, photometric systems given a 
detailed understanding of the filter characteristics.  After calculating 
$k$-corrections with \texttt{kcorrect}, we remove any mean $k$-corrected 
color versus redshift trends with a low order polynomial to account for the 
passive evolution of stellar populations between observed epochs. This 
evolutionary correction is normalized to a redshift of 0.3, near the median 
of the LRG redshift distribution.  The color evolution correction is small 
(less than 2\% for all bands in both samples) and does not affect any of 
the results presented here.  We  further assume a galaxy dims by one 
magnitude per $\Delta z=1$ change in redshift in the $g$-band due to the 
passive evolution of its stellar populations.  Again, this correction is 
normalized to $z=0.3$.  In the remainder of this paper, all of the color
and luminosity measurements refer to quantities which have been adjusted 
for the $k$-correction and evolution.

\begin{deluxetable}{cc}
\tablecolumns{2}
\tablenum{1}
\tablewidth{0pt}
\tablecaption{Effective Wavelengths of Shifted Bandpasses}
\tablehead{
  \colhead{Filter} & 
  \colhead{$\lambda_{\mbox{eff}}$}\\
\colhead{(1)} &
\colhead{(2)} }
\startdata
$^{0.16}g$  & 4026 \\
$^{0.16}r$  & 5307 \\
$^{0.16}i$  & 6441 \\
$^{0.16}z$  & 7687 \\
$^{0.37}g$ &  3408 \\
$^{0.37}r$ &  4494 \\
$^{0.37}i$ &  5454 \\
$^{0.37}z$ &  6509 \\
\enddata
\end{deluxetable}

Throughout this paper, we work in the  $^{0.16}g$, $^{0.16}r$, $^{0.16}i$, 
$^{0.16}z$ and $^{0.37}g$, $^{0.37}r$, $^{0.37}i$, $^{0.37}z$ systems for 
the low- and moderate-redshift galaxy samples, respectively.   Here, the 
notation $^{z_0}g$ indicates the reported AB magnitudes are derived through
a standard $g$ filter that has been blueshifted by $z_0$.  For a 
filter system which has been shifted by $z_0$, the $k$-correction for galaxies 
at that redshift is trivial, $-2.5 \, \mbox{log} \, (1+z_0)$, and is 
independent of the galaxy spectral energy distribution.  In choosing 
shifts which nearly match the median redshift of each sample, we minimize 
the uncertainty introduced through the $k$-corrections \citep{blanton2003b}.   
Table 1 lists the effective wavelengths for each of the bandpasses used here.

\subsection{Sample Construction}

In order to investigate the color-magnitude relation of massive galaxies in 
the range $z \sim 0.1-0.4$, we create two samples from the SDSS spectroscopic 
database to be analyzed separately; Figure \ref{sample} shows the selection 
cuts used to define each of these samples. From the MAIN sample of galaxies, 
we extracted 16622 galaxies in the redshift range $0.1<z<0.2$ satisfying the 
rest-frame color cut 
\begin{equation}
^{0.16}(g-i) > 8.374 + 0.32 \left( M_g - 5\, \hbox{log}\, h \right)
\end{equation}
which approximates a cut of constant stellar mass \citep{BdJ2000}.  This 
selection criterion results in a cut diagonally across the red sequence, 
which could bias our measurements of the scatter in  red sequence colors 
and will greatly affect any slope measurement of the color-magnitude relation. 
In order to avoid biased slope measurements and test the consistency of our  
scatter measurements, we also construct a sample of 5100 galaxies in the 
luminosity range $-22.5<M_g - 5 \, \hbox{log} \, h<-21.5$ with 
$^{0.16}(g-i) > 1.46$ for comparison (shown by the dashed region in 
Figure \ref{sample}).

We further select 2782 LRG galaxies from the SDSS Southern Survey 
in the redshift range $0.3<z<0.4$ which satisfy
\begin{equation}
^{0.37}(r-z) > 7.769 + 0.32 \left( M_g - 5\, \hbox{log}\, h \right)
\end{equation}
and have been imaged 6 or more times.   Equation 3  is the same rest-frame 
color cut applied to the low-redshift galaxies (Equation 2) shifted to 
the $^{0.37}(r-z)$ color --- the effective wavelengths of the filters 
that define each of these colors are similar, and thus these colors probe 
similar regions in the rest-frame galaxy spectrum (see Table 1).   Of 
these 2782 sample galaxies, 62\% were imaged more than 9 times while 20\% 
of the galaxies were imaged more than 12 times in the SDSS Southern Survey.   
As in the low-redshift sample, we adopt a simple constant color cut, 
$^{0.37}(r-z) > 0.825$ and $-22.5<M_g - 5 \, \hbox{log} \, h<-21.5$,  
to define a  subsample of 1380 high-redshift galaxies to test the 
consistency of our scatter measurements and allow us to determine 
the slope of the color-magnitude relation without a strong cut across 
the red sequence.

\begin{figure}[ht]
\centering{\includegraphics[angle=0, width=3in]{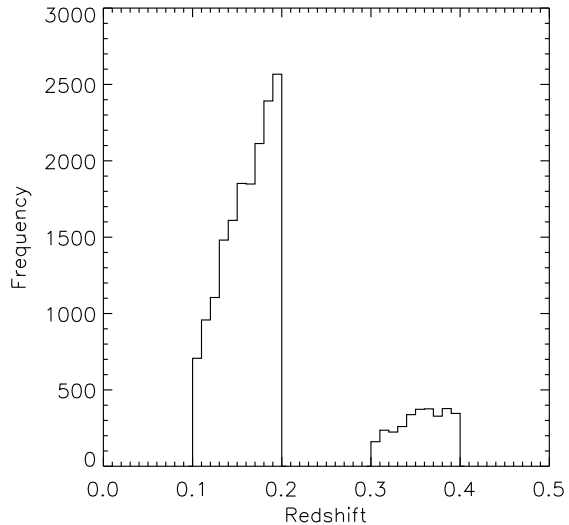}}
\caption{\scriptsize Redshift distribution of the galaxies selected for 
this work is shown.  The low-redshift sample was drawn from the MAIN SDSS 
galaxy sample while the moderate-redshift sample is composed of  LRG galaxies
selected from the SDSS Southern Survey.  The low-redshift sample has robust
photometry in one observation epoch  as the galaxies are upwards of five 
magnitudes brighter than the SDSS detection limit while the higher redshift 
galaxies, which extended two magnitudes fainter, have been imaged repeatedly 
allowing for more precise photometric measurements.}
\label{zhist}
\end{figure}

We find no differences within our quoted error between the dispersion 
around the CMR 
measured from galaxies defined using the stellar mass cut and the constant 
color cut; throughout the remainder of this paper, scatter measurements 
will be based upon galaxies drawn in the first manner while all slope 
measurements are referenced to samples created using the latter method. 
Figure \ref{zhist} shows the redshift distribution of the galaxies used 
to measure the scatter around the color-magnitude relation in this paper.

In detail, the red sequence is populated by elliptical and S0 galaxies as 
well as late-type galaxies reddened by dust. The arrow in the lower right 
of Figure \ref{sample} shows the reddening vector given by the \citet{odonnell}
 extinction curve. In order for a dusty
late-type galaxy to scatter into our sample, it must have a large unobscured
luminosity as we only consider the most luminous red-sequence galaxies in 
this work.  There is a clear paucity of very luminous blue galaxies in
Figure \ref{sample} and thus the contamination from dusty spiral galaxies is 
likely small in our sample. Furthermore, \citet{eisenstein2003}  constructed 
the average spectrum of massive galaxies, such
as those used in our analysis, and found that only 5\% of the 
galaxy spectra in the $-22.5<M_r<-22$ luminosity range 
were contaminated by  emission lines uncharacteristic of early-type galaxies.
In the remainder of this paper, we will use the
red-sequence and early-type classifications synonymously.
  
Throughout this paper, we will refer to the galaxies considered here as 
field galaxies as they are not specifically selected to reside in clustered 
environments.  It is important to note, however, that galaxies which are of 
similar color and luminosity as those in our sample tend to lie in over-dense 
environments \citep{hogg2003}.  \citet{zehavi2005} found the clustering 
strengths and mean separations of LRGs are comparable to those of poor 
clusters or rich groups.  For comparison, of the LRGs in the volume surveyed 
by \citet{bahcall2003} to find galaxy clusters with 13 or more detected 
cluster members, 16\% are projected less than 500 kpc from the inferred 
location of a cluster and within 0.05 of the photometric redshift for 
that cluster.  A detailed examination of the density dependence of the scatter 
around the CMR is conducted in \S5.

\pagebreak

\begin{figure}[ht]
\centering{\includegraphics[angle=0, width=3.5in]{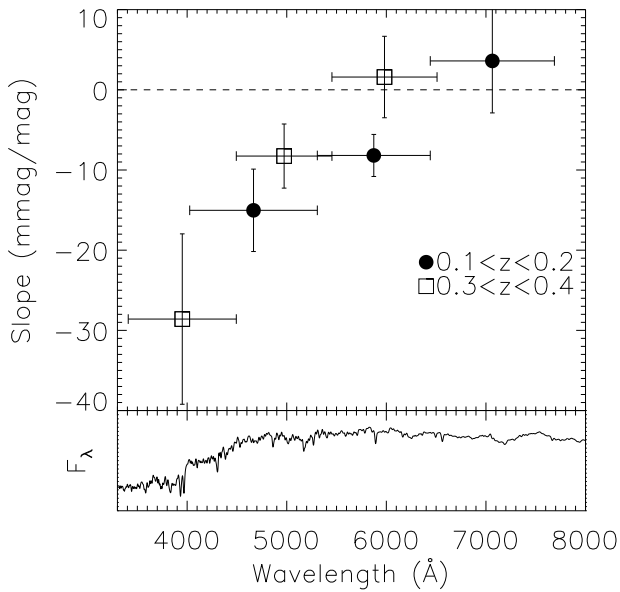}}
\caption{\scriptsize Measured slope of the CMR as a function of the rest-frame
bandpass.  The endpoints of the horizontal error bars mark the effective
wavelengths of the filters used to define the bandpass while the vertical
error bars show the measured uncertainty on each slope measurement.  The 
spectrum of an early-type galaxy is plotted as a reference. The two symbols
represents the two galaxy samples - the hollow squares mark measurements on 
the moderate-redshift ($0.3<z<0.4$) sample while the filled circles show the 
data from low-redshift (0.1 $<z<$ 0.2) galaxies. }
\label{fig:slopeplot}
\end{figure}

\section{The Color-Magnitude Relation}

When determining the slope of the color-magnitude relation, we only consider 
the galaxies selected with our simple luminosity and constant color cuts as 
any slope measurements based on samples with a strong cut across the red 
sequence will be quite biased. The galaxy samples considered in this work 
only span a small range of luminosities and thus are not ideally suited for 
measurements of the slope of the CMR -- we simply use our measurement of 
this quantity to remove the mean relationship before determining the scatter 
around it.  The slopes and zero-points for the CMR for several colors are 
listed in Table 2.  Figure \ref{fig:slopeplot} shows the measured slopes 
of the CMR as a function of the effective wavelengths of the two bandpasses 
used to define the color.  The slope in bluer colors is more pronounced than 
that for redder bands which tend to show small or negligible slopes. This 
trend is consistent with a metallicity sequence as the 
primary driver along the 
red-sequence.  As noted by \citet{GY} using models from \citet{K97},
the flattening of the color-magnitude relation 
toward redder colors can also be 
reproduced by an age trend along the red sequence but only if a very specific 
relationship between galaxy mass and star formation
time holds with very little scatter.  In the 
bluest colors, which have been used historically in the literature, the 
slopes measured here are in general agreement with past work; for comparison, 
\citet{hogg2004} report a slope of $-0.022$ mag mag$^{-1}$ in the 
$^{0.1}(g-r)$ color for a large sample of galaxies at $z\sim0.1$.

\begin{deluxetable*}{rcrrcrrcr}[hb]
\tablecolumns{9}
\tablenum{2}
\tablewidth{0pt}
\tablecaption{Slope of the Color-Magnitude Relation}
\tablehead{
  \colhead{} & 
  \multicolumn{2}{c}{Total Sample} & \colhead{} &  
  \multicolumn{2}{c}{Field Sample} & \colhead{} &
  \multicolumn{2}{c}{Cluster Sample} \\
  \cline{2-3} \cline{5-6} \cline{8-9} \\
   \colhead{Color} &
  \colhead{Zero-point\tablenotemark{a}} &
  \colhead{Slope} & \colhead{} & 
  \colhead{Zero-point\tablenotemark{a}} &
  \colhead{Slope} & \colhead{} &
  \colhead{Zero-point\tablenotemark{a}} &
  \colhead{Slope}  \\
\colhead{}&
\colhead{mag} &
\colhead{mmag/mag} & \colhead{} & 
\colhead{mag}  & 
\colhead{mmag/mag} & \colhead{} & 
\colhead{mag} &
\colhead{mmag/mag}\\
\colhead{(1)} &
\colhead{(2)} &
\colhead{(3)} & \colhead{} &
\colhead{(4)} &
\colhead{(5)} & \colhead{} &
\colhead{(6)} & 
\colhead{(7)}}
\startdata
$^{0.16}(g-r)$ & 1.162 & $-15.6 \pm  5.2$ & & 1.162  & $-10.5 \pm 6.7$   & & 1.164 &  $  -12.4 \pm  7.8$ \\
$^{0.16}(r-i)$ & 0.449 & $-8.10 \pm  2.6$ & & 0.449  & $-6.7  \pm 3.4$   & & 0.451 &  $ -7.6  \pm  4.3$ \\
$^{0.16}(i-z)$ & 0.329 & $3.50 \pm  6.4$  & & 0.329  & $ 4.2  \pm 8.2$   & & 0.329 &  $  5.5  \pm  10.4$ \\
$^{0.16}(g-i)$ & 1.611 & $-26.1 \pm  5.5$ & & 1.610  & $-21.5  \pm 7.4$  & & 1.612 &  $-22.9  \pm  8.2$ \\
$^{0.37}(g-r)$ & 1.750 & $-23.3 \pm  12.1$ & & 1.750 & $-15.5  \pm 16.1$ & & 1.754 &  $-10.6  \pm  19.5$ \\
$^{0.37}(r-i)$ & 0.613 & $-3.8 \pm   4.3$  & & 0.612 & $ 0.41 \pm  6.2$ & & 0.614  &  $-3.6  \pm  8.4$ \\
$^{0.37}(i-z)$ & 0.379 & $ 5.0 \pm   6.1$  & & 0.377 & $ 0.58 \pm  7.7$ & & 0.381  &  $ 7.21 \pm  9.9$ \\
$^{0.37}(r-z)$ & 0.991 & $ 1.8 \pm   8.5$  & & 0.980 & $ 3.4  \pm  3.5$ & &  0.983 &  $8.1  \pm  1.4$ \\
\enddata
\tablenotetext{a}{The zero-point is defined at $M_g-\logh=-21.8$.}
\end{deluxetable*}

\begin{figure}[hb]
\centering{\includegraphics[angle=90, width=3.5in, viewport=100 50 450 750, clip]{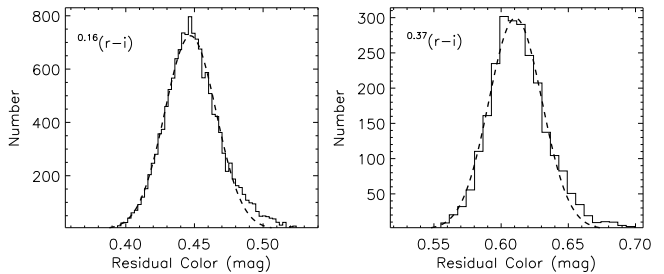}}
\caption{ \scriptsize Scatter around the color-magnitude relation for both 
galaxy samples.  The $^{0.16}(r-i)$ and $^{0.37}(r-i)$ residual colors, 
normalized to the characteristic color of a galaxy with $M_g-\logh=-21.8$ 
are shown for the low and intermediate redshift galaxies respectively.  Note
that the dispersion of this relation is quite small.  The dashed line is the
best-fit Gaussian to the data.  Table 3 lists the measured scatter around 
the CMR for all of the colors measured in this paper. }
\label{fig:sigmaresid}
\end{figure}

Figure \ref{fig:sigmaresid} illustrates the typical scatter about the 
mean color-magnitude relation for the  $^{0.16}(r-i)$ and $^{0.37}(r-i)$ 
colors; the measured dispersion about this relationship is quite small.  
The second column of Table 3 lists the measured dispersion around the CMR  
for each of the colors used in this study.  For each color studied, we fit 
a Gaussian to the observed color distribution using Poisson errors on each 
color bin and define the dispersion of the best-fit Gaussian to be the 
scatter in galaxy colors on the red sequence.  By fitting the 
histograms rather than calculating the standard deviation directly, 
we place more weight on the 
core of the distribution compared to the wings.  Thus, 
our scatter measurements will not be heavily affected by
any  non-red-sequence 
interlopers in the wings of the color distributions.  
As this method could potentially depend on the 
binning used, we are careful to vary the bin size for each fit and ensure the 
measured scatter is not dependent on our choice of bin size.  It should be 
noted that past work has used the standard deviation to quantify the scatter 
in galaxy colors.  While the standard deviation and Gaussian widths may give 
systematically different dispersions in the case of non-Gaussian 
distributions, Figure \ref{fig:sigmaresid} shows the color distribution on 
the red sequence is well described by a Gaussian and thus the two measurements 
are comparable.    Bootstrap resampling of the data \citep[e.g.][]{press} 
was employed to determine the uncertainty on our scatter measurements.  It 
is important to note that the scatter measurements in the second column of 
Table 3 represent upper limits to the intrinsic scatter in galaxy colors as 
these values are not corrected for systematic or experimental  effects, 
which could be significant.  

In order to investigate the level of skewness of the observed color 
distribution, we define $\beta_{n}$ to be the absolute value of the 
color difference between the 
$n\%$ quantile and the median of the distribution.  The level of asymmetry 
in the distribution is then estimated using
\begin{equation}
\gamma = \frac{\beta_{10} - \beta_{90}}{\beta_{10}+\beta_{90}} .
\end{equation}
For this statistic, a maximally skewed distribution would have $\gamma=\pm1$ 
such that a tail of galaxies to the red would result in $\gamma<0$. A symmetric
distribution would have $\gamma=0$.  We adopt
this approach rather than a traditional third moment 
calculation as the calculation
of higher moments of a distribution is strongly dependent on the wings of 
the distribution where contamination may be important.  
For the histograms shown in Figure \ref{fig:sigmaresid}, we find that 
$\gamma = -0.027 \pm 0.019$ for the $^{0.16}(r-i)$ color and $\gamma = -0.064 
\pm 0.048$ for the $^{0.37}(r-i)$ color.   Both of these measurements
reflect the slight over-abundance of red galaxies compared to a 
Gaussian in both panels of   
Figure \ref{fig:sigmaresid}.  This red tail is 
due to the diagonal cut across the red sequence imposed by our stellar
mass selction criterion;  red galaxies are selected to 
lower luminosities than blue 
galaxies.  Since the luminosity function of galaxies is steeply rising 
toward lower luminosities in this range, more red galaxies are selected 
than blue galaxies. For comparison, we measure $\gamma = 0.001 \pm 0.029$ 
for the $^{0.16}(r-i)$ color and $\gamma = -0.005 \pm 0.057$ for 
the $^{0.37}(r-i)$ color when the simple luminosity cuts are used 
to select our sample galaxies.


\begin{figure}[!ht]                                                          
\centering{ \includegraphics[angle=0, width=3in, viewport=50 50 500 700, clip]{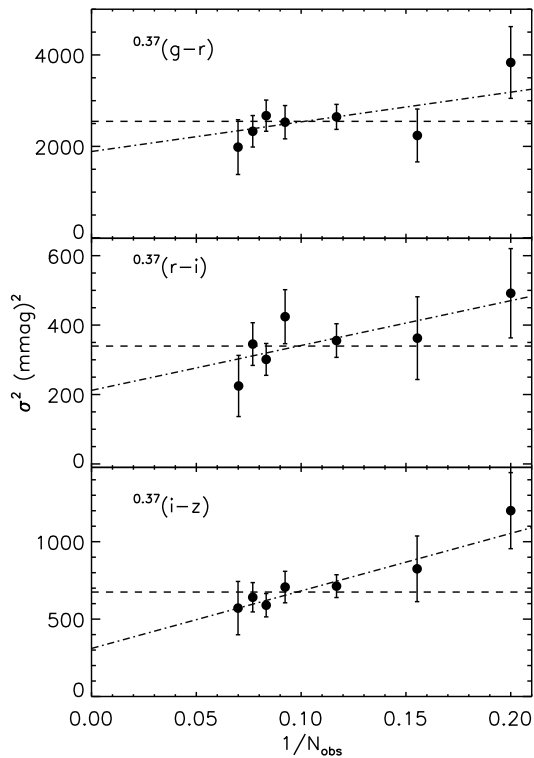}}                     
\caption{\scriptsize Relationship between the number of observations and the 
measured scatter around the color-magnitude relation.  The dotted line 
shows the weighted mean of the data while the dot-dashed line marks the best
fit line to the data. The data for each color are not statistically 
consistent with a constant value but are well described by a line, as is 
expected.   We extrapolate the best fit line to infinite measurements to 
find the intrinsic scatter in galaxy colors; these values are listed in the
fourth column of  Table 3.}
\label{fig:richardson}

\end{figure}


Since our intermediate redshift sample is drawn from regions that have been 
repeatedly imaged by the Sloan Digital Sky Survey, we can assess the level to
which the dispersion we observe in the color-magnitude relation is dependent 
on measurement noise.  If the scatter is contaminated by measurement noise, 
we would expect the squared scatter to decrease inversely with the number of
observations :
\begin{equation}
\sigma_{\rm obs}^2 = \sigma_{\rm intrinsic}^2 + \frac{A}{N_{\rm obs}} \, \, .
\end{equation}
Figure \ref{fig:richardson} shows the results of this test; the methods used
for the ensemble sample were used in the same manner to calculate the scatter
and error for  each bin. In each bin of $N$-measurements, we include galaxies
which were imaged a total of $N$ times, and thus each bin contains a unique 
subsample of galaxies. For the moderate-redshift sample, the third and fourth
columns in Table 3 lists the  instrumental contribution and intrinsic galaxy
scatter predicted by extrapolating the fits to Equation 2 to infinite 
observations for each of the three colors.

\begin{deluxetable*}{cccccc}[hb]
\tablecolumns{6}
\tablenum{3}
\tablewidth{0pt}
\tablecaption{Scatter Around the Color-Magnitude Relation}
\tablehead{
\colhead{Color} & 
\colhead{Measured Scatter} &
\colhead{Instrumental Scatter} &
\colhead{Intrinsic Scatter} &
\colhead{Field\tablenotemark{a}} &
\colhead{Clustered\tablenotemark{a}} \\
\colhead{} &
\colhead{mmag} &
\colhead{mmag} &
\colhead{mmag} &
\colhead{mmag} &
\colhead{mmag} \\
\colhead{(1)} &
\colhead{(2)} &
\colhead{(3)} &
\colhead{(4)} &
\colhead{(5)} &
\colhead{(6)} } 
\startdata
$^{0.16}(g-r)$ & $39.6\pm0.4$ & $17.6\pm3.6$ & $35.4\pm3.7$
  &  $40.5\pm0.4$  &  $35.9\pm0.8$ \\ 
$^{0.16}(r-i)$ & $18.7\pm0.2$ & $15.1\pm3.9$ & $11.0\pm4.0$ 
 &  $19.1\pm0.2$  &  $17.3\pm0.3$ \\ 
$^{0.16}(i-z)$ & $26.6\pm0.4$ & $20.9\pm4.2$ & $20.9\pm4.2$ 
 &  $26.7\pm0.5$  &  $23.4\pm0.8$ \\
$^{0.16}(g-i)$ & $40.0\pm0.4$ & $21.2\pm4.1$ & $34.4\pm4.1$
 &  $42.0\pm0.5$ & $36.1\pm0.7 $ \\
$^{0.37}(g-r)$ & $48.0\pm1.2$ & $20.4\pm6.1$ & $43.5\pm6.2$ 
 &  $48.8\pm1.4$  &  $47.3\pm2.2$ \\
$^{0.37}(r-i)$ & $19.7\pm0.5$ & $12.3\pm3.1$ & $14.6\pm3.1$ 
 &  $20.1\pm0.6$  &  $18.9\pm1.0$ \\
$^{0.37}(i-z)$ & $27.9\pm0.7$ & $21.6\pm4.5$ & $17.6\pm4.6$ 
 &  $29.2\pm0.8$  &  $24.4\pm1.4$ \\
$^{0.37}(r-z)$ & $35.9\pm1.1$ & $18.4\pm5.2$ & $30.8\pm5.2$
 &  $38.1\pm1.2$ & $28.4\pm2.1$ \\
\enddata
\tablenotetext{a}{Field and clustered values are {\it measured} 
scatter, not the intrinsic scatter in the galaxy colors.}
\end{deluxetable*}

While the intrinsic dispersion around the color-magnitude relation at moderate
redshift can be extracted from the measured scatter based on subsamples of 
galaxies with different number of measurements, the low-redshift galaxies do
not allow for this test as most galaxies in this sample have been imaged 
only once.  In order to estimate the intrinsic scatter in galaxy colors for
this sample, we identified 5839 of our low-redshift galaxies that have been
imaged more than once under photometric conditions.  We adopt the average 
root-mean squared variation of each photometric measurement as the instrumental
error for each color of interest.   We find the mean measurement uncertainty
to be 17.6, 15.1, and 20.9 mmag in the $^{0.16}(g-r)$, $^{0.16}(r-i)$, 
and $^{0.16}(i-z)$ bands, respectively.  The internal dispersion in each color
is the difference, in quadrature, between the measured scatter and inferred
mean uncertainty in our single pass photometry. For the low-redshift 
galaxies, the third and fourth columns of Table 3 lists the observational 
error and intrinsic scatter determined by this correction.  While our 
estimate of the instrumental contribution to the observed color dispersion 
corrects for uncorrelated errors on each independent measurement, we do not 
correct for systematic errors common to all observations which could cause 
an overestimation of  the intrinsic dispersion in galaxy colors.

\begin{figure}[ht]

\centering{\includegraphics[angle=0, width=3.5in]{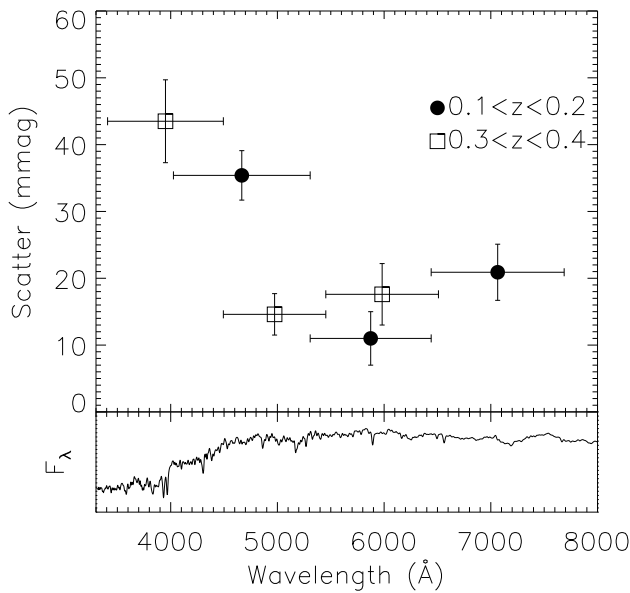}}
\caption{ \scriptsize Intrinsic scatter in the color-magnitude relation as 
a function of the observed bandpass.  The error bars and symbols are defined
as in Figure \ref{fig:slopeplot}.  The spectrum of an early-type galaxy
is shown for comparison.    The scatter in redder bands tends to be smaller
than in blue bands.  This is not unexpected, as small changes in a galaxy's
age, metallicity, or dust attenuation affect the blue portion 
of a its spectrum more than the
red. }
\label{fig:sigmaplot}
\end{figure}

Figure \ref{fig:sigmaplot} shows our scatter measurements as a function of 
the effective wavelength of the filters  used to construct each color. The 
scatter about the CMR is clearly wavelength dependent; the bluest colors, 
which are most affected by small changes in the age, metallicity, or dust 
attenuation of the 
system, have systematically larger dispersions than the reddest colors where 
changes in metallicity and age have less impact. 

\section{Star Formation History of Massive Galaxies}

Here, we consider three  simple star formation histories and 
derive the range of permitted parameters for these histories
following a method similar to that of \citet{vandokkum1998}. In these toy 
models, we will assume that only age variations between red-sequence galaxies 
creates the scatter in early-type galaxy colors.  In reality, the scatter
across the CMR is likely driven by some combination of age, metallicity,
dust, and possibly $\alpha$-enhancement variations across the red-sequence.
These models will thus illustrate the broadest range of 
ages allowed by our measurements; the inclusion of any variation in the other
galaxy properties will only constrain the age more strongly.  

In this section, we will primarily consider the high-redshift sample, as 
these are the intrinsically youngest galaxies in our sample and hence 
should provide the tightest 
constraints on the past star formation history of massive early-type galaxies. 
An identical analysis of the low-redshift sample provides results consistent
with those presented below.  In all three model cases, the low-redshift 
galaxies provide looser constraints than the high-redshift sample; star
formation is allowed to continue to slightly later times.

First, we consider a star formation history composed of a single 
delta-function burst.  The probability of a burst occurring is distributed 
uniformly between the earliest allowed time for galaxy formation, $t_0$, 
set to $z=10$, and $t_{\rm max}$, the time at which the youngest early-type 
galaxies form.   Using a grid of Monte-Carlo simulations with 5000 galaxies 
per realization, we create samples of galaxy spectra synthesized using the 
method of \citet{bc2003} with solar metallicity, a \citet{salp} initial mass 
function, and varying values of $t_{\rm max}$.  We find that realizations 
allowing galaxies younger than 8.8 Gyr ($z \sim 1.4$) generate color 
distributions with standard deviations larger than that measured.  

We perform a similar test using a scheme in which 80\% of the stars in a 
galaxy are formed in a single burst at $z=10$ and 20\% are formed in a second 
burst which occurs at a random time between $z=10$ and a minimum age, as done 
by \citet{vandokkum1998}.  For this scenario, the secondary bursts must occur 
between $10>z>0.9$ to reconstruct the observed scatter in the color-magnitude 
relation.  For bursts that generate less than  5\% of a galaxy's stars, 
the time scale for the secondary burst is unconstrained - small amounts of 
star formation can occur at late epochs without violating the constraints 
placed by our measurements.


\begin{figure*}[ht]

\centering{\includegraphics[width=5in]{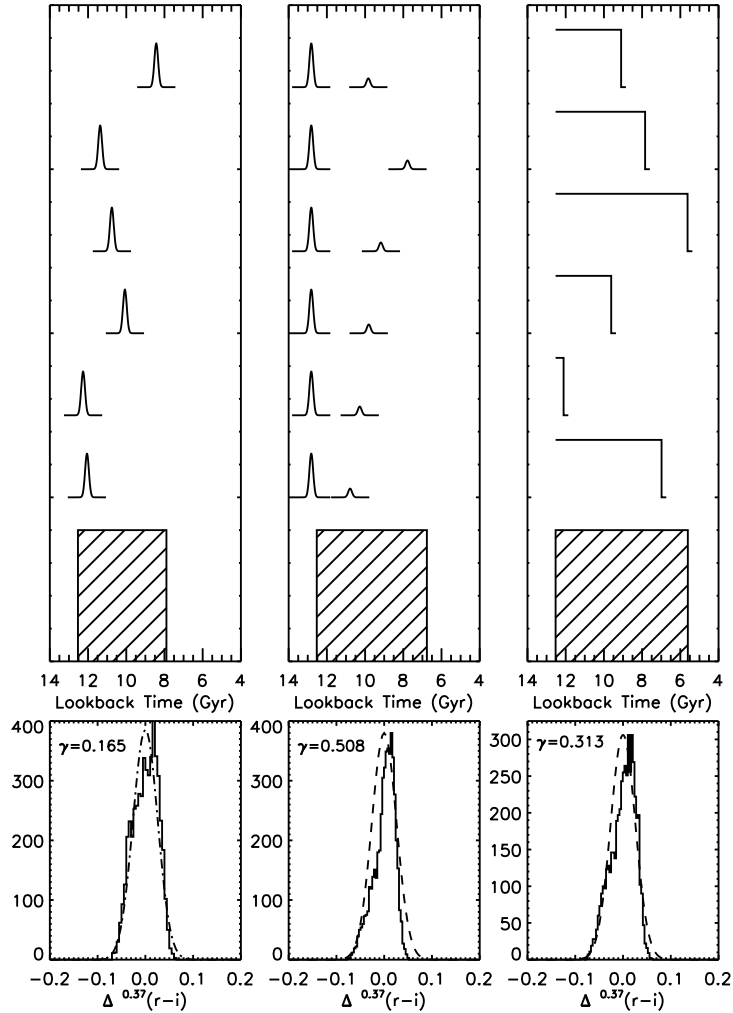}}

\caption{ \scriptsize Reproduction of  the observed scatter in rest-frame 
color of early-type galaxies using Monte-Carlo simulations following
a method similar to that of \citet{vandokkum1998}. For each 
simulation, the top panel shows several examples of star formation 
histories drawn from the simulation.  The shaded region denotes the 
allowed epoch of star formation for that simulation.  Each of the bottom
panels shows the distribution of residual colors around the color-magnitude
relation for the realization that most closely matches the observed scatter. 
The $\gamma$ parameter, as defined in the text, is listed for each of the
distributions.  For comparison with the
 color distributions in Figure \ref{fig:sigmaresid},
we have added typical observational errors to the color distributions.  
The dashed line in each panel shows a Gaussian with the same mean and 
dispersion as  the simulated colors. The left-most panel illustrates the 
allowed epoch of star formation for a SFH composed of a single delta function 
burst.  For galaxy formation beginning at a redshift of 10, the youngest 
early-type galaxies must have formed by $z\sim1.13$.  The middle panel 
illustrates the allowed range of times for a secondary burst which contributes 
20\% of the stars to the galaxy.  This secondary burst must have occurred 
before a $z\sim0.84$.  Finally, for a constant star formation history, all 
star formation must have ended by $z\sim0.62$ to reproduce the observed 
scatter in the color-magnitude diagram at $z=0.37$. For reference, the 
lookback time at $z=0.37$ is 3.89 Gyr.  A comparison with Figure 
\ref{fig:sigmaresid} shows that the color distributions predicted by these 
models do not match the observed colors of massive galaxies, a sign that 
the models used here are overly simple.  This is echoed by the mismatch 
between the $\gamma$ value measure from the data and those 
measured for each of the model distributions.  }
\label{fig:models}
\end{figure*}

Finally, we create a grid of models in which star formation begins at $z=10$ 
and continues at a constant rate until it is truncated at $t_{\rm f}$.  
For each realization, we allow $t_{\rm f}$ to range from $z=10$ to a 
critical cutoff time at which all galaxies have stopped forming stars.   
Star formation in this manner must have ended by $z \sim 0.8$ in order 
to reproduce the observed scatter in the color-magnitude relation at $z=0.37$.

The allowed range of parameters and the distribution of galaxy colors 
found in each model are shown graphically in Figure \ref{fig:models}.    
To allow comparison with Figure \ref{fig:sigmaresid}, typical observational 
errors have been included in the color distributions in the lower panels 
of the figure.  Interestingly, all of these models predict similar scatter
trends to those seen in Figure \ref{fig:sigmaplot}.  In all three cases, the 
scatter in the $^{0.37}(g-r)$ color is nearly double that predicted for the 
$^{0.37}(r-i)$ and $^{0.37}(i-z)$ colors.
While each of the three models succeeds in reproducing 
the observed scatter in early-type galaxy colors, none of the models 
adequately recreates the color distribution measured from real galaxies, 
a sign that the models chosen do not perfectly track the true evolution of 
the star formation history of early-type galaxies. The $\gamma$ parameter, 
as defined in Equation 4, for each of the model distributions is shown in 
Figure \ref{fig:models}. The three models produce color distributions with
$\gamma=(0.165, 0.508, 0.313)$ compared to the value of $-0.064$ calculated
for the observed sample of galaxies at the same redshift. 
The difference between 
the observed skewness and that determined from our simple models 
reiterates the mismatch between the observed and predicted color 
distributions. In our simulations, we assume all galaxies have the same 
metallicity and no dust attenuation. Also, the onset of star formation is 
uniform in the second two models.  In reality, these parameters are not 
constant, and thus the sharp truncations seen on the red edges of 
model color distributions will be blurred by the addition of
other complications to the models.   However, any of these corrections will
increase the observed scatter in galaxy colors thus  driving the last epoch 
of star formation needed to satisfy our constraints to earlier times.

\section{Comparison with Clustered Environments}


A majority of the work on the color-magnitude relation of early-type galaxies 
has focused on cluster galaxies at various redshifts.  While some work 
\citep[e.g.][]{sv1978,ltc1980,kbb1999} tackles the problem in the field, 
these studies suffer from small numbers of galaxies.   The measurement of the 
scatter around the CMR presented here, on the other hand, focuses on the most
massive field galaxies at moderate redshifts using a statistically 
significant sample.  We can now, for the first time, statistically compare 
the dispersion around the CMR in both clustered and field environments.  

In order to explore the environmental dependence of the color-magnitude 
relation within our data set, we use SDSS imaging of 5305 deg$^2$ to 
identify a comparison sample of 16 million normal galaxies down to $r=21$. 
At the location of each LRG galaxy in our sample, we count the number of 
neighboring galaxies within a $1 \, h^{-1}$ Mpc (proper) aperture which have 
colors similar to those expected for an $L^*$ galaxy on the red sequence, 
and luminosities in the $M^*-0.5$ to $M^*+0.4$ range; the details of the 
luminosity and color cuts are given in \citet{eisenstein2005}. 

Using the number of neighbor galaxies as a proxy for the environment of each
LRG in our sample, we divide each of our galaxy sets into a clustered
subsample and a low density subsample. We define LRG galaxies with more
than 5 (7) neighboring galaxies in the low- (moderate-) redshift samples 
to be in clustered environments while a field subsample is comprised of 
galaxies with fewer neighbors.  This threshold was chosen such that the 
clustered sample in both redshift bins composes 30\% of the total galaxy
sample at that redshift.  To place these values in physical context, 
at $z=0.16$ ($z=0.37$) there are 25.3 (170) tracer galaxies per square 
degree.  Given the angular diameter distance of 398 (712) $h^{-1}$ Mpc to 
that redshift, this yields a total of 1.65 (3.46) background galaxies per
aperture on average.  The average number of neighbors in the field sample 
is  2.3 (3.9) galaxies while the clustered galaxies have a mean of   9.1 
(11.4) neighbors.  Thus our clustered samples are over-dense by a factor of 
4 (3)  compared to the field galaxies.  

\citet{hogg2004} found that the slope and zeropoint of the color-magnitude 
relation do not depend strongly on the local galaxy environment.  
We find the same result; Table 2 
lists the measurements of the slope and zeropoint of the relationship for 
the ensemble sample as well as for the field and clustered subsamples. The
variation between the two samples is small.  In the following calculations, 
we use the ensemble slope and zeropoint measurements as these are measured 
with the highest signal-to-noise.  The results are unaffected if the 
individual measurements of the slope and zeropoint are used for 
each subsample.

The fifth and sixth columns of Table 3 list the measured scatter in galaxy 
colors for the field and clustered galaxies respectively.  In all of the  
bands, the scatter for the field subsample is larger than that in clustered 
environments.  If we assume the scatter introduced by instrumental effects 
is the same for both the field and clustered subsamples, the same correction 
determined for the ensemble galaxy sample can be applied to these subsamples,
as well.  With this correction, we find that the scatter in galaxy colors in 
dense environments is $11\pm2\%$ smaller than for that of isolated systems in
both redshift intervals studied here. 

Our clustered galaxies
reside in a range of densities, not simply rich clusters, as have often
been used for similar studies, and thus we likely underestimate the
significance of the difference in scatter between the field and strongly
clustered environments. Furthermore, our field sample is not completely
composed of isolated galaxies as the average number of neighbor galaxies
in this sample is larger than the average density of background galaxies.
Again, this leads to an underestimation of the true difference in the
dispersion around the color-magnitude relation for field and clustered
galaxies.

The larger scatter in field galaxy colors supports previous claims of an 
environmental dependence for the scatter around the color-magnitude relation 
for early-type galaxies \citep[e.g.][]{ltc1980,kbb1999},
 though the effect measured here is weak.  Comparisons
of the scatter in galaxy colors for clustered galaxies often draw on galaxies
from a single cluster whereas our clustered sample is created by combining 
galaxies from a number of clustered environments.  It is possible that the
star formation histories of galaxies in a single cluster are more highly
coordinated than the star formation histories between different clusters,
thus the scatter measured using galaxies from a single cluster would be 
smaller than when several clusters are combined.  As the fraction of blue
galaxies is larger in the field compared to clustered environments, the field
subsample may have a larger contamination fraction compared the the clustered
subsample. 
This would also cause an increased scatter in field galaxies colors 
compared to the galaxies in more dense environments.


\begin{figure*}[ht]

\centering{\includegraphics[angle=90, width=7in]{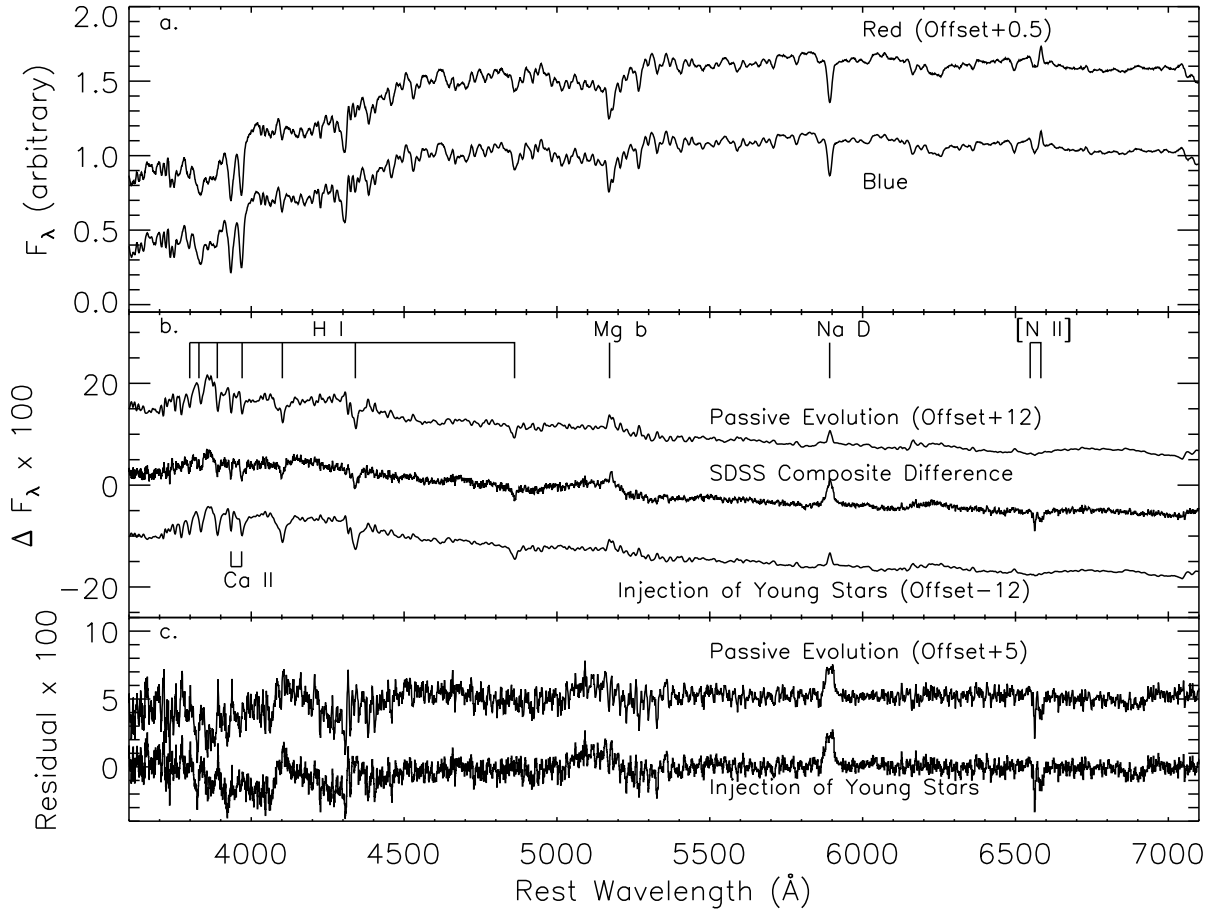}}

\caption{ \scriptsize (a) Composite spectra of galaxies between 1-$\sigma$ 
and 3-$\sigma$  of the mean $^{0.16}(g-r)$ color.  The gross properties of 
the composite red and blue spectra are quite similar. (b) The difference 
spectrum between the blue and red galaxies (middle) compared to the difference 
spectrum predicted for a passively evolving population (top) and an old 
population with a recent injection of young stars (bottom).  The difference 
spectrum shows clear signs of enhanced Balmer absorption lines in the blue 
galaxies as well as stronger molecular features in the red galaxies.  The 
red galaxies have stronger sodium absorption, leading to the ``emission" 
feature in the difference spectrum.  (c) Residuals between the data and 
comparison synthesis spectra.  The general properties of the observed 
difference spectrum are matched in both comparisons, though the model 
consisting of an old galaxy with a small amount ($\sim 3 \%$) of star 
formation in the past 1 Gyr slightly over-predicts the continuum strength 
in between 3900 $\hbox{\AA}$ and 4100 $\hbox{\AA}$ compared to the 
passively evolving model.  }
\label{fig:coadd}
\end{figure*}

\section{Average Galaxy Spectrum Across the Red Sequence}

Having measured the scatter in galaxy colors, we next investigate the spectral 
differences between galaxies by examining the average  spectrum of massive 
early-type galaxies across the red sequence.  The composite early-type 
galaxy spectrum has been investigated as a function of environment, 
luminosity, redshift, size, and velocity dispersions \citep{eisenstein2003,
bernardi2003a}; here, we construct the composite spectrum for red sequence
galaxies of different mean colors.  Since the galaxies in our low-redshift 
sample are brighter, and thus have higher quality spectra, and are more
numerous than the higher redshift galaxies, we only consider these galaxies
in our analysis. 

To construct the average spectrum of the galaxies in each color bin, we 
first shift the spectra according to the observed redshift  and then normalize
the observed SDSS spectrum by the flux at rest-frame 5000 $\hbox{\AA}$.  
We augment the SDSS pixel mask to include all pixels within 10 $\hbox{\AA}$
of the bright sky lines at 5577, 5890, 5896, 6300, 6364, 7245, and 7283 
$\hbox{\AA}$.  Since the galaxies of interest are in the redshift range 
$0.1<z<0.2$, the masked pixels occur at different rest-frame wavelengths 
for each galaxy, leaving the mean spectrum unaffected. We then find the 
number-weighted mean of the spectra, ignoring all masked pixels in the mean.  

Galaxies are divided into two distinct color bins -- red galaxies having
$^{0.16}(g-r)$ colors between 1-$\sigma$ and 3-$\sigma$ redward of the mean
and a blue sample in the same range blueward of the mean.  The red galaxy 
bin contains 1321 galaxies with a mean $^{0.16}(g-r)$ color of 1.20 and 
$M_g - 5 \, \hbox{log} \, h = -21.3$ while the blue sample contains 1161 
galaxies with mean color and luminosity of 1.09 and 
$M_g - 5 \,  \hbox{log} \, h = -21.5$.   Figure \ref{fig:coadd}(a) shows 
the coadded spectra of the red and blue galaxies. The gross properties of each 
composite spectrum are similar. Figure \ref{fig:coadd}(b) shows the 
difference between the blue and red spectra.  This difference spectrum 
shows clear signs of enhanced Balmer absorption in the blue galaxies 
compared to the red composite, indicating that, on average,
the blue galaxies contain 
more young stars,  in agreement 
with past studies \citep[e.g.][]{trager2000,bernardi2005}. 
 While the red composite shows stronger 
[N~$\!${\footnotesize II}] emission lines relative to the blue composite, 
no other prominent emission lines, such as [O~$\!${\footnotesize II}], 
[O~$\!${\footnotesize III}], and H$\alpha$ are present; this emission line 
difference is likely due to a larger fraction of AGN in the red galaxies 
compared to the blue.

We consider two possible evolutionary scenarios to explain the spectral 
differences between the blue and red galaxies on the red sequence: pure age
(i.e. passive) evolution of a single age stellar population and a recent 
epoch of star formation mixed with an existing old population.  In the first
scenario, we fit each composite spectrum with a solar metallicity, single
age, stellar synthesis model created using the method of \citet{bc2003}. 
The red and blue spectra are best fit by 6.3 Gyr and 3.7 Gyr populations, 
respectively.  In the second scenario, we find the blue composite spectrum 
is well fit by a 6.3 Gyr old stellar population mixed with a 1 Gyr population
containing 3.5\% of the stellar mass and we model the red composite with a 
simple 6.3 Gyr population. The difference spectra for both scenarios are 
shown in Figure \ref{fig:coadd}(b).  While each of the two model hypotheses 
would suggest quite different evolutionary histories for massive early-type
galaxies, the predicted difference spectra are remarkably similar.  The
model consisting of a recent burst of star formation slightly over-estimates 
the continuum in the difference spectrum near 3900 $\hbox{\AA}$ and thus 
provides a marginally poorer fit to the data than the passive evolution model.

The residuals between the data and the model difference spectra are shown in 
Figure \ref{fig:coadd}(c).  Both of the models fail to match the strengths 
of the  $\hbox{CN}_2$ and Mg $b$ features near 4140 and 5160 $\hbox{\AA}$.  
\citet{eisenstein2003} interpret these differences between observations and 
synthetic spectra as a enhanced $\alpha$-to-iron ratio present in the 
galaxies but not included in the synthetic models. Also, several weak 
iron features are evident in the two model spectra 
near 4300 and 5300 $\hbox{\AA}$ which are missing in the observed difference
spectrum.  In a pure-age model for the evolution across the red sequence,
the enhanced blue continuum provided by the
hotter stars in the young population  
dilutes the metal lines seen in the older population.  This dilution
leads to differences in the strengths of the metal lines of the two 
population models, thus causing the residual iron features in
both model difference
spectra.  The absence of these residuals in our observed difference spectrum 
indicates that a pure-age model for the spectral differences between red and 
blue galaxies on the red sequence is an incomplete model. 
In order to match 
the variations  of the spectral features across the red 
sequence, a more detailed model is required.  This model should include both 
age and metallicity variation and, more importantly, non-solar abundance 
ratios.  Until 
spectral synthesis models with realistic $\alpha$-enhancement 
prescriptions are available, more quantitative analysis 
of our observed difference spectrum 
is difficult as the interplay between age, metallicity, and $\alpha$-to-iron 
ratio variations must be included for a full understanding of the physical 
differences between these galaxies.

Another clear difference 
between the models and the data is the strong Na D feature near 5892 
$\hbox{\AA}$.  Here, the red composite spectrum has considerably more Na D 
absorption than predicted by the stellar synthesis models.  As the model 
spectra do not contain absorption due to the interstellar medium, this 
discrepancy may indicate that the red galaxies contain more neutral gas than 
their blue counterparts. 

\section{Conclusions}

We present new multi-color observations of a large sample of galaxies in the 
range $0.1<z<0.4$ gathered from the Sloan Digital Sky Survey. These galaxies 
represent the most massive galaxies in the Universe  and are not selected to 
be in clustered environments. Past studies of the color-magnitude relationship
have focused on $L^{*}$ galaxies in clusters, thus our work focuses on an area 
of parameter space previously unexplored.  Using the robust photometry of 
bright galaxies provided by the SDSS as well as multiple observations of 
moderate-redshift galaxies, we are able to construct the color-magnitude 
relations for massive field galaxies in several colors.  While our data only 
span one magnitude in luminosity, and thus are not ideal for in-depth analysis 
of the slope and zeropoint of the CMR, our large sample of 
galaxies allows us to measure the dispersion about this relationship 
quite well.  The scatter around the red sequence is 20 mmag in the 
reddest bands while the bluest colors have nearly double this value.

We construct several toy model star formation histories in order to place 
constraints on the evolutionary processes at work in massive galaxies.  We
find that the majority of star formation in these early-type galaxies must 
have occurred before $z \sim 1$, in agreement with other authors.   For 
secondary bursts of star formation with strengths less than 5\% of the total 
stellar mass of the system, our data are consistent with bursts of star 
formation at any epoch.  However, one should note that none of the toy models 
reproduce the observed distribution of galaxy colors.

Furthermore, we make comparisons between our new measurement of the scatter 
around the color-magnitude relation for massive field galaxies and early-type
galaxies in clusters.  An internal comparison shows that the scatter around
the CMR in dense environments is $11 \pm 2 \%$ smaller than that measured 
using more isolated galaxies. 

The composite spectra of LRG galaxies redward and blueward of the mean galaxy
color are physically distinct. The blue galaxies show stronger Balmer 
absorption than the red galaxies, consistent with the presence of a younger
stellar population. Simple models slightly favor passive evolution of stellar 
populations as the cause of the observed differences in spectra across the 
red sequence compared to a more recent injection of young stars onto an 
old stellar population, though the difference between the predicted spectra
from these models is small.  

\section{Acknowledgments}
We would like to thank Michael Blanton, Scott Burles, Douglas Finkbeiner,
David Hogg, David Schlegel, John Moustakas, Amelia Stutz, and Jane Rigby
for software and helpful comments during the preparation of this paper. 
We further thank the anonymous referee for a thorough and critical reading
of our paper which lead to substantial improvements to the text. 
This research made use of the NASA Astrophysics Data System.   RJC is 
funded through a National Science Foundation Graduate Research Fellowship. 
DJE is supported by NSF grant AST 339020 and an Alfred P. Sloan Research 
Fellowship. 

Funding for the creation and distribution of the SDSS Archive has been
provided by the Alfred P. Sloan Foundation, the Participating Institutions, 
the National Aeronautics and Space Administration, the National Science
Foundation, the U.S. Department of Energy, the Japanese Monbukagakusho, 
and the Max Planck Society. The SDSS Web site is http://www.sdss.org/.

The SDSS is managed by the Astrophysical Research Consortium (ARC) for 
the Participating Institutions. The Participating Institutions are The 
University of Chicago, Fermilab, the Institute for Advanced Study, the Japan 
Participation Group, The Johns Hopkins University, the Korean Scientist Group,
Los Alamos National Laboratory, the Max-Planck-Institute for Astronomy 
(MPIA), the Max-Planck-Institute for Astrophysics (MPA), New Mexico State 
University, University of Pittsburgh, University of Portsmouth, Princeton 
University, the United States Naval Observatory, and the University of 
Washington.

\clearpage


\end{document}